\documentclass[12pt,a4paper]{article}
\usepackage{graphics}
\usepackage{epsfig}
\usepackage[cp1251]{inputenc}
\usepackage[T2A]{fontenc}
\usepackage[english]{babel}
\usepackage{cite}

\author{A.~P.~Kuznetsov$^{1,2}$, A.~V.~Savin$^{1,2}$ and
D.~V.~Savin$^1$\\[14pt]
\it $^1$Departament of nonlinear processes, Saratov State University,\\
\it 83 Astrakhanskaya str., Saratov, 410012, Russia\\[14pt]
\it $^2$Institute of Radio Engineering and Electronics\\
\it of Russian Academy of Science, Saratov branch,\\
\it 38 Zelenaya str., Saratov, 410019, Russia\\
\\
E-mail: dmitry\_new@rambler.ru}
\title{On some properties of nearly conservative dynamics of Ikeda
 map and its relation with the conservative case}
\date{}

\begin{document}
\DeclareGraphicsExtensions{.jpg,.pdf,.mps,.png}
\maketitle
\begin{abstract}
The behavior of the well-known Ikeda map with very weak
dissipation (so called nearly conservative case) is investigated.
The changes in the bifurcation structure of the parameter plane
while decreasing the dissipation are revealed. It is shown that
when the dissipation is very weak the system demonstrates an
"intermediate" type of dynamics combining the peculiarities of
conservative and dissipative dynamics. The correspondence between
the trajectories in the phase space in conservative case and the
transformations of the set of initial conditions in the nearly
conservative case is revealed. The dramatic increase of number of
coexisting low-period attractors and the extraordinary growth of
the transient time while the dissipation decreases have been
revealed. The method of plotting a bifurcation trees for the set
of initial conditions has been used to classify existing
attractors by it's structure. Also it was shown that most of
coexisting attractors are destroyed by rather small external
noise, and the transient time in noisy driven systems increases
still more. The new method of two-parameter analysis of
conservative systems was proposed.
\end{abstract}
\section{Introduction}

It is well known that the behavior of conservative and dissipative
systems differs essentially. E.g., the majority of nonlinear
conservative systems can demonstrate the chaotic dynamics
practically at all values of parameters, but usually it realizes
in very small area of phase space. On the other hand dissipative
system demonstrates chaotic behavior only at certain values of
parameter, but the basin of that chaotic attractor usually
occupies a considerable area in phase space (see e.g. \cite{1,2}).
Furthermore, differences in dynamics lead to significant
difference in numerical methods for investigation. So, practically
all methods applicable for dissipative systems are based on the analysis
of the attractors, while conservative systems have no attractor at
all. As a result two practically independent branches studying
 conservative and dissipative systems correspondingly had been
formed in nonlinear dynamics.

But physically dissipative and conservative systems are not
isolated and for a big number of systems a transition from
dissipative to conservative systems while continuous change of the
parameters can occur. In this case dynamics changes smoothly from
dissipative to conservative and some "intermediate" behavior
should occur "near" the conservative case. Investigation of this
process seems to be very interesting because such behavior should
demonstrate both conservative and dissipative features. Such
investigations were began in \cite{3} for so-called rotor map, or
standard map, which conservative modification is the classical
model of conservative system (see e.g. \cite{2}). It had been
revealed, that the rotor map could demonstrate very peculiar
dynamics combining some features of conservative and dissipative
dynamics while its Jacobian approaches 1. In particular, a huge
number of co-existing low-period periodic attractors can be
observed, which leads to significant dependence of the dynamics on
the initial conditions. We should note that such dependence is
typical for conservative systems.

In this paper we try to investigate the dynamics of another
classical model - the Ikeda map - while dissipation decreases and
the system evolutes from dissipative to conservative.

\section{Ikeda map}
The Ikeda map
\begin{equation}
z_{n+1}=A+Bz_nexp(i(|z_n|^2+\psi))
\end{equation}
 had been proposed by Ikeda {\itshape et
al.} \cite{4} to describe the dynamics of light in the ring
cavity. Now it is one of the classical models of nonlinear
dynamics demonstrating a big number of it's basic phenomena. We
would like to emphasize that the Ikeda map is an approximate
stroboscopic map for a driven nonlinear oscillator \cite{5} and so
can roughly describe a big amount of systems of different nature.
Let's consider the connection between map (1) and the pulse driven
nonlinear oscillator
\begin{equation}
\ddot{x}+\gamma\dot{x}+{\omega_0}^2 x+\beta x^3=\sum C\delta(t-nT)
\end{equation}
more detail. The Ikeda map can be obtained from the Eq. (2) if one
solves the autonomic equation between
 external pulses by method of slow amplitudes. The connection between parameters of Eqs.
(1) and (2) is given by following formulae:
\begin{equation}
A=\frac{C}{\omega_0}\sqrt{{\frac{3\beta}{8\omega_0}}\frac{1-e^{-\gamma
T }}{\gamma}}, B=e^{-\gamma T/2}, \psi=\omega_0 T.
\end{equation}

The Jacobian of this map is equal to $B^2$ , hence $B=1$
corresponds to the conservative system, $B<1$ - to dissipative and
$B\approx 1$ - weakly dissipative (nearly conservative) system.
Nowadays the dynamics of the Ikeda map in dissipative case is well
studied (see e.g. \cite{5,6,7}). In particular, it is known that
so called "crossroad area" structures \cite{8,9} typical for
driven nonlinear oscillator exist in the parameter plane of this
map (fig.~1~a).

At the transition to weakly dissipative case the general structure
of the parameter plane remains practically the same, but some
changes occur (see fig.~1~b). For example, the "crossroad area"
structure changes, and the degenerate flip \cite{10} point
appears on the period-doubling line, indicating the appearance of
the supercritical period-doubling bifurcation. Also it should be
mentioned that a transition to chaos occurs now at smaller values
of parameter $A$.
\begin{figure}[h]
\centering
\includegraphics[width=16cm,keepaspectratio]{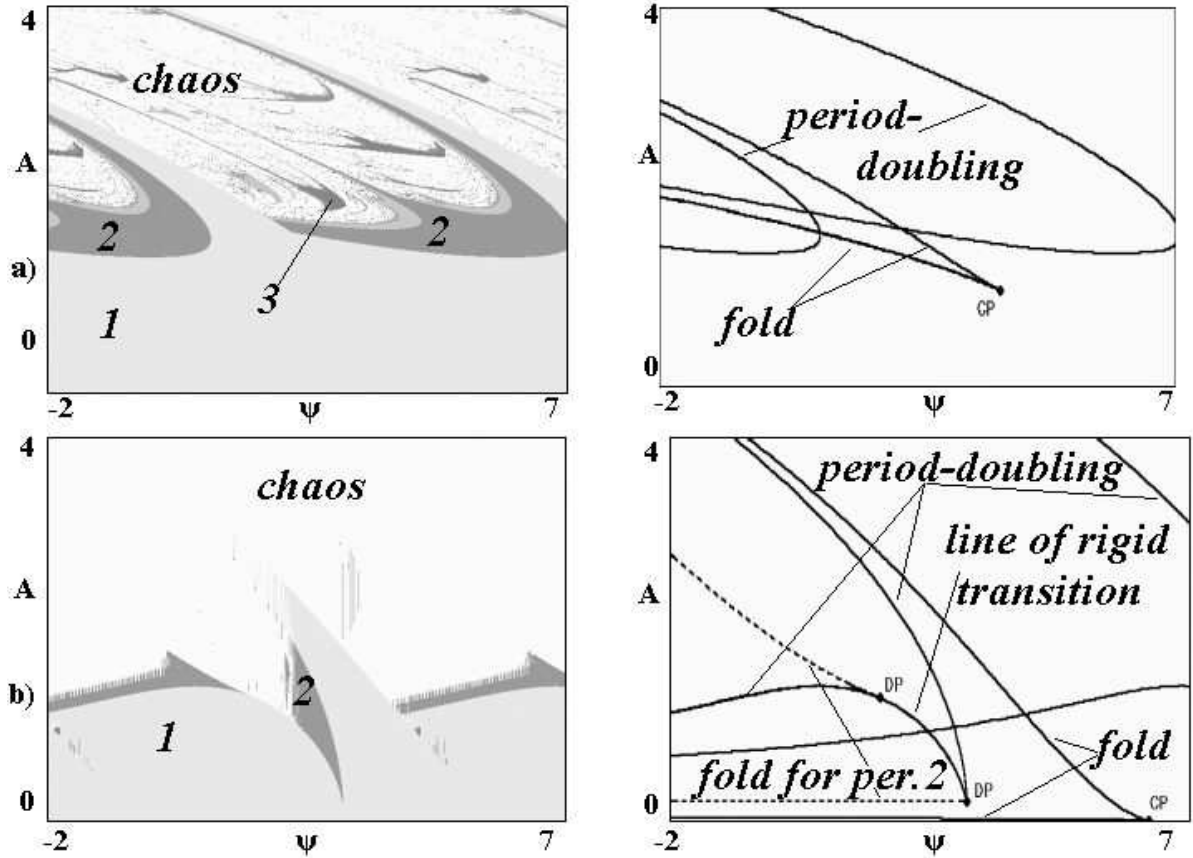}
\caption{The structure of the parameter plane of the Ikeda map (1)
with essential (a, $B=0.3$) and weak (b, $B=0.99$) dissipation. On
the left side there are so called charts of dynamical regimes
where the stability regions of the cycles of different periods are
shaded with different gray shades. On the right side there are
bifurcation lines of Ikeda map (1) plotted by the {\it Content}
program. CP means a cusp point, DP --- the degenerated flip point.}
\end{figure}

\section{Evolution of attractors with the decrease of dissipation}
Now let's turn to the analysis of the phase space structure in the
nearly conservative case. For this we have taken "cloud" of points
in the phase space and have observed the consecutive stages of its
condensing on the attractor (see fig.~2). We can see that at the
first stages of cloud evolution (fig.~2~a) structures similar to
the phase portrait in conservative case (fig.~2~d) arise. At
further stages of the dynamics several focuses which attract other
points can be seen (see fig.~2~b, where these focuses are marked
by stars). The location of those focuses corresponds to
the location of the elliptic fixed points in conservative case. It
shows us that on small times of evolution a weakly dissipative
system demonstrates nearly conservative dynamics with further
transition to dissipative dynamics.
\begin{figure}[h]
\centering
\includegraphics[width=8cm,keepaspectratio]{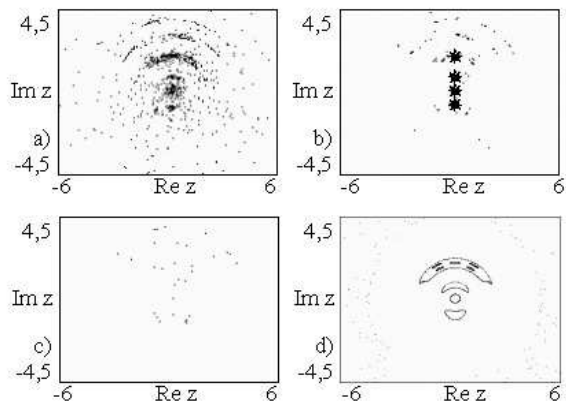}
\caption{The different stages of the evolution of cloud of initial
conditions for the Ikeda map (1) in weakly dissipative case (a--c) and
phase portrait in the conservative case (d). The figures a--c differs by
the number of missed iterations: a) 200; b) 660; c) 6000. The
locations of focuses are marked by stars at fig.~b. Values of
parameters: $A=0.5; \psi=3\pi/4; B=0.99$ (a-c), $B=1$ (d).}
\end{figure}

When all transients have died away, several attracting points can
be seen in the phase space (see fig.~2~c) although this point on
the parameter plane is inside the region of the period 1 at the
fig.~1~b. This allows us to suppose that several periodic
attractors coexist at this point of the parameter plane, i.e.
multistability exists.

For investigation of this phenomenon in weakly dissipative case a
method of drawing bifurcation trees for a set of initial
conditions on one diagram had been proposed in \cite{3}. At each
value of the control parameter one should take a set of the
initial conditions, make a sufficient number of iterations to cut
off all transients and then plot several consequent iterations at
the "parameter - variable" plane. Such diagrams allow us to obtain
the number of coexisting attractors and to trace their
transformations while changing the parameter. It seems natural to
choose a set of initial conditions in the domain where an
attractor exists to decrease the amount of calculations. From (1)
we can obtain $|z_{k+1}|\leq A+B|z_k|$.  It is obvious that an
attractor can't exist in the domain, where $|z_{k+1}|\leq |z_k|$
because there $|z_k|$ decreases. The boundary of this domain we
can determine from the condition $|z_{k+1}|\leq A+B|z_k|\leq
|z_k|$. Hence, $|z_k|\geq A/(1-B)$ is the domain where an
attractor can't exist. Therefore, the domain of attractor
existence is bounded by the condition $|z|\leq A/(1-B)$ , i.e.
$|z|_{max}=A/(1-B)$. We'll take initial conditions on the mesh in
the rectangle $[-x_{max}, x_{max}] \times [-y_{max}, y_{max}]$,
where $x_{max}=y_{max}=A/(1-B)$.

Bifurcation diagrams for different values of dissipation parameter
B plotted by this method are shown in fig.~3. On all of them we
can see the "basic" attractor, which arises at $A=0$ and
demonstrates the first period-doubling bifurcation at values $A$
near 1. This attractor has the largest basin so usually it is
represented on the charts (fig.~1). Besides it there are some
"secondary" attractors. They arise at non-zero values of parameter
$A$ and demonstrate classical transition to chaos by
period-doubling cascade. Their bifurcation trees have rather
simple "classical" structure. Let's refer such attractors as the
attractors of first type.
\begin{figure*}
\centering
\includegraphics[width=16cm,keepaspectratio]{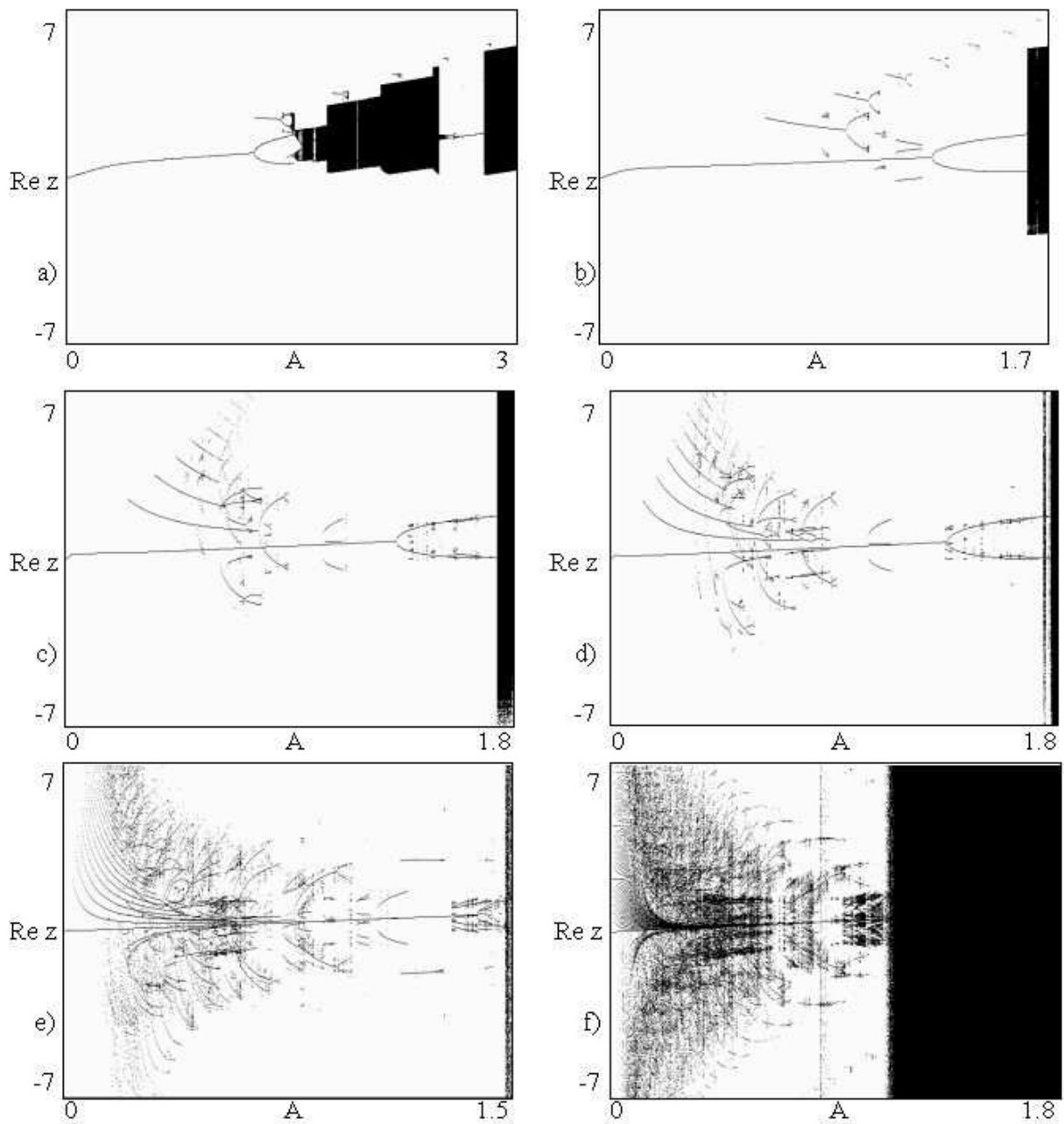}
\caption{Bifurcation diagrams for the map (1) for different values
of dissipation parameter: a) $B=0.5$; b) $B=0.75$; c) $B=0.9$; d)
$B=0.95$; e) $B=0.99$; f) $B=0.999$. $\psi=0$.}
\end{figure*}

At relatively strong dissipation ($B=0.5$) a number of such
"secondary" attractors is not very big but it increases while
decreasing of dissipation and they arise at smaller values of $A$.
Also the distance between them along the  $A$ axe decreases.
Furthermore, fragments with essentially more complex dynamics
arise on bifurcation diagrams.

Attractors corresponding to these trees arise at rather large
values of $A$ and are characterized with a smaller interval of
their existence on $A$ axe than attractors of the first type.We
shall refer them as the attractors of the second type. The number
of such attractor also increases while decreasing the dissipation.

Now let's investigate a weakly dissipative case (fig.~3~f,e) in
more detail. First we consider the case $B=0.99$ (fig.~3~f). It
should be marked that the transient time becomes extremely long
(up to 500000 iterations) at this case and more than ten times
exceeds the transient time for essentially dissipative system. In
the fig.~4 the bifurcation diagrams for different transient time
are shown. It can be seen that besides it's extreme length, the
transient time essentially depends on the parameter $A$. This
dependence is extremely irregular so the areas where transient
time is less then 20000 iterations interchange with areas where it
is more than 200000 iterations.
\begin{figure}[h]
\centering
\includegraphics[width=16cm,keepaspectratio]{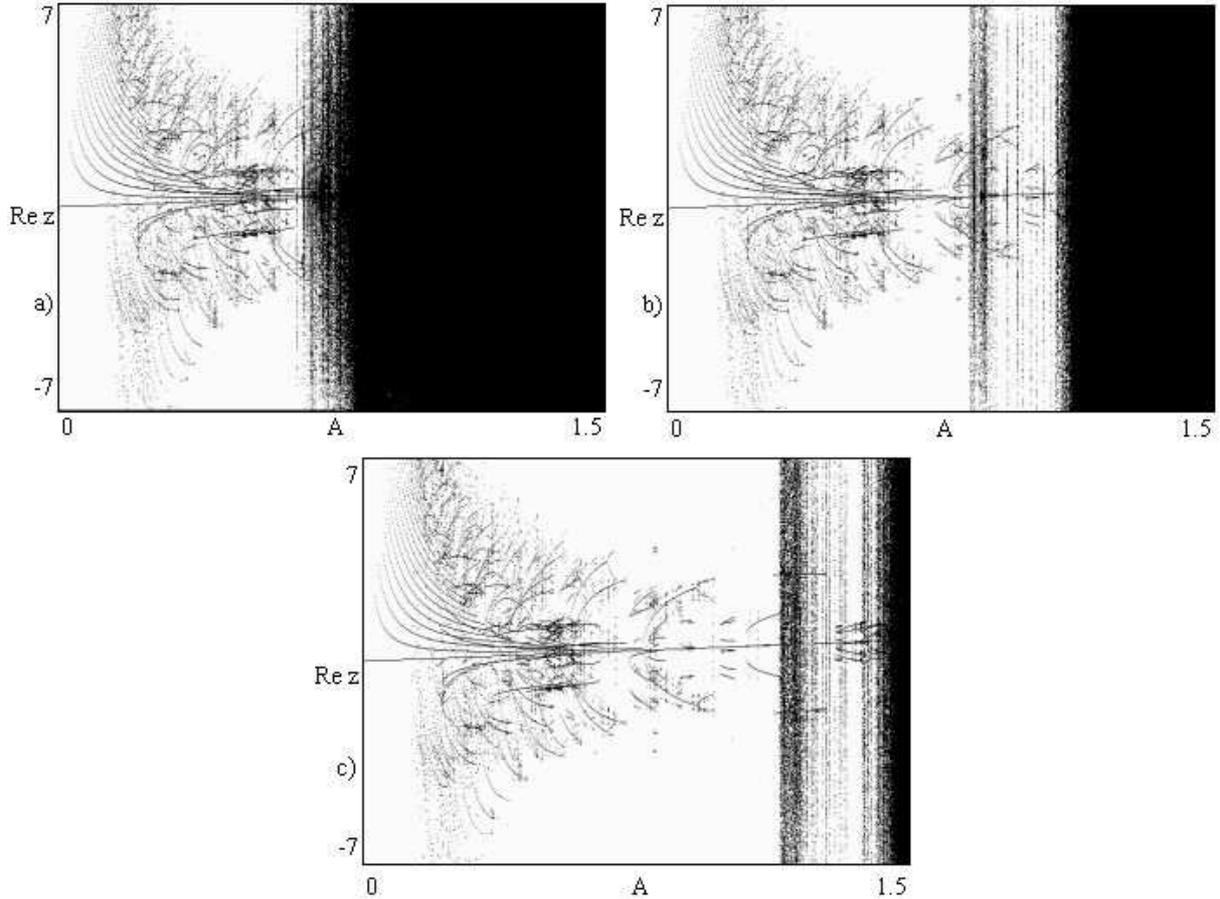}
\caption{Bifurcation diagram for the map (1) with a different
number of missed iterations (transient process): a) 5000; b)
20000; c) 200000. Parameters $B=0.99; \psi=0$.}
\end{figure}

Now let us discuss the diagram structure when all transients have
died away (fig.~3~f). It demonstrates a big number of attractors
both of the first and second types. It should be noted that in
fact there exist a considerably greater number of attractors than
it can be seen on the bifurcation diagram because many attractors
have the basin smaller than a period of the initial conditions
mesh. It is confirmed by the fact that the structure of the
bifurcation diagram complicates essentially when a number of
points of the mesh increases from 400 to 10000 points (in
particular, a number of attractors of the second type increases).
At the same time there are no changes constrained with the
attractors of the first type, which shows that they have larger
basins and, consequently, their observation in realistic system is
more probable.

There are no chaotic attractors on the bifurcation diagram for the
mesh with 400 initial conditions. We think that it is also because
their basins are too small. The period-doubling cascade for the
majority of attractors is observed only up to period 2 which can
be caused by two reasons. First is that in conservative systems
the distance between two consecutive period-doubling points
decreases much faster than in dissipative (corresponding constant
$\delta$=8,7210972… is essentially greater than well-known
Feigenbaum constant 4,6692016...), so the regions of high periods
can't be represented on the bifurcation diagram. The second is
that the attractors can undergo a crisis reaching the boundary of
it's basin. This assumption seems to be more realistic; some
arguments in its favor will be discussed in section 4.

Besides of the bifurcation diagrams for variable $x=Re\ z$
diagrams for other variables such as $y=Im\ z$ and $|z|^2$ has been
built (fig.~5). On $|z|^2$ diagram (fig.~5~b,c) it is clearly seen
that the attractors of the second type are attractors of period 2
and higher because corresponding bifurcation trees always consists
of two and more branches. Also it should be noted that while the
$x$ value for the attractors of the first type increases with the
decrease of the  parameter $A$, the $y$ value decreases, and
$|z|^2$ value remains practically constant. It means the point in
the phase plane moves on a circle, approaching the real axe while
$A$ tends to zero.
\begin{figure}[h]
\includegraphics[width=16cm,keepaspectratio]{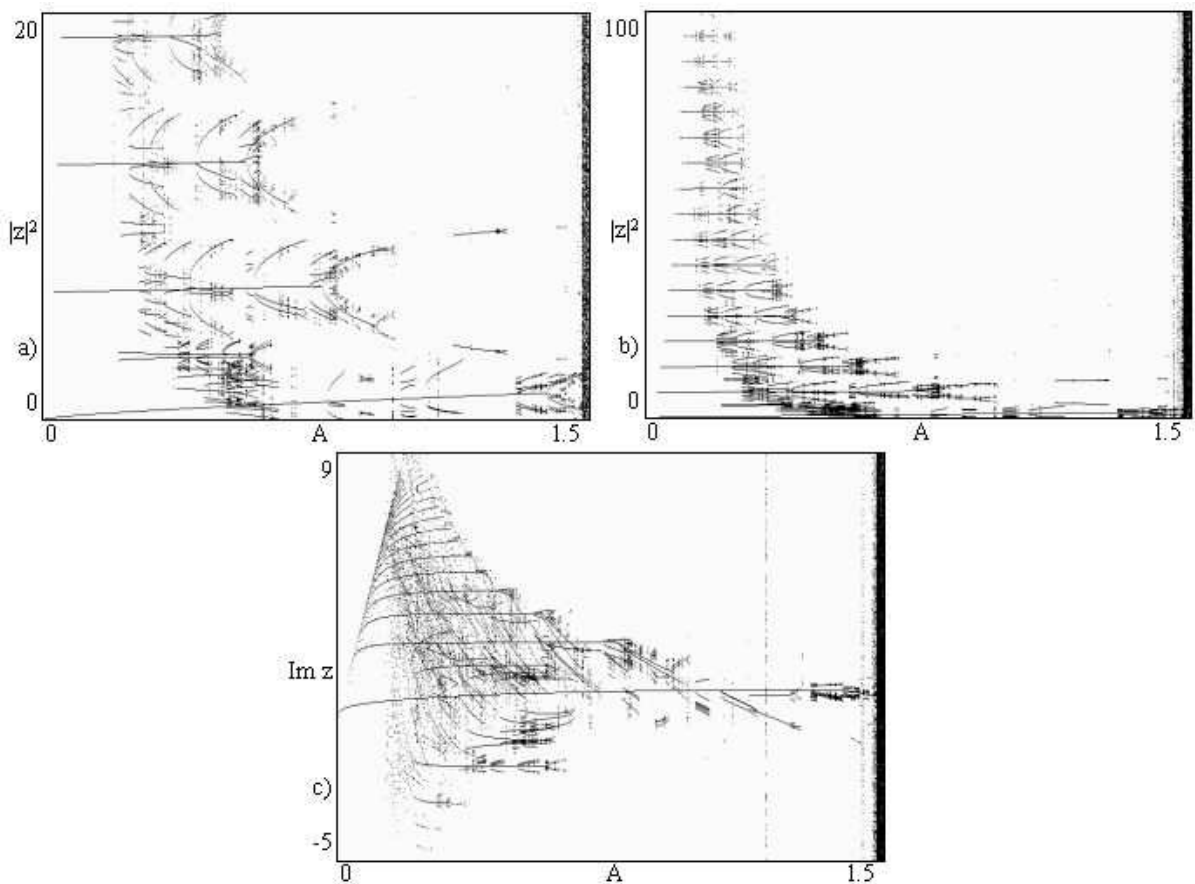}
\caption{Bifurcation diagrams for variables $y=Im\ z$ and $|z|^2$.
The parameters $B=0.99; \psi=0$.}
\end{figure}

For weaker dissipation ($B=0.999$, fig.~3~f) transient time
reaches 5000000 iterations and a number of coexisting attractors
extends extremely, but evidently there are no qualitative changes
in the structure of the diagram, e.g. all attractors can be
divided into the same two types.

In this case the attractors of the first type form several
"families" which tends to the horizontal lines with the increase
of the parameter $A$. We can say that in previous case there are
only one "family" which includes all attractors of the first type
and in this case there are several "families". The "center" of
each "family" is the attractor  (stable fixed point) with very
weak dependence on the parameter represented by horizontal line on
the diagram. Empty lines between the "families" may be seen on the
diagram so it is naturally to suppose that also some unstable
fixed points with very weak dependence on the parameter exists
being the boundaries between the "families" of the attractors.

\section{Noisy driven weakly dissipative Ikeda map}

In the previous sections we have shown that the Ikeda map
demonstrates an exceptional variety of coexisting low-period
periodic attractors in the case of weak dissipation. But it seems
significant to explore how the dynamics of the system will change
with adding an external noise, because it always exists in real
systems. Let's consider noisy driven system as follows:
\begin{equation}\label{2}
z_{n+1}=A+Bz_nexp(i(|z_n|^2+\psi))+\varepsilon \xi_n
\end{equation}
 where $\xi_n$ is
a random real value (uniformly distributed on the segment [-1;1]
in our numerical experiments) and $\varepsilon$ can be interpreted
as an amplitude of noise. It should be noted that in this form the
system can describe the nonlinear oscillator driven by external
pulses with fixed intervals but random amplitude.

In noisy driven system the transient time becomes even more longer
and approaches 700000 for $B=0.99$. In the fig.~6~a,b bifurcation
diagrams for different amplitudes of noise (a, b) are shown. In
the fig.~6~c they are laid one on another to compare it's
structure. It is well seen that a large amount of attractors (and
the larger the amplitude of noise is, the larger is this amount),
and between them the majority of the attractors of the second
type, is destroyed by noise influence. The destruction of the
attractors can be explained by the fact that their basins are too
small and a noise influence simply "throw" the point out of the
basin.
\begin{figure}[h]
     \leavevmode
\includegraphics[width=16cm,keepaspectratio]{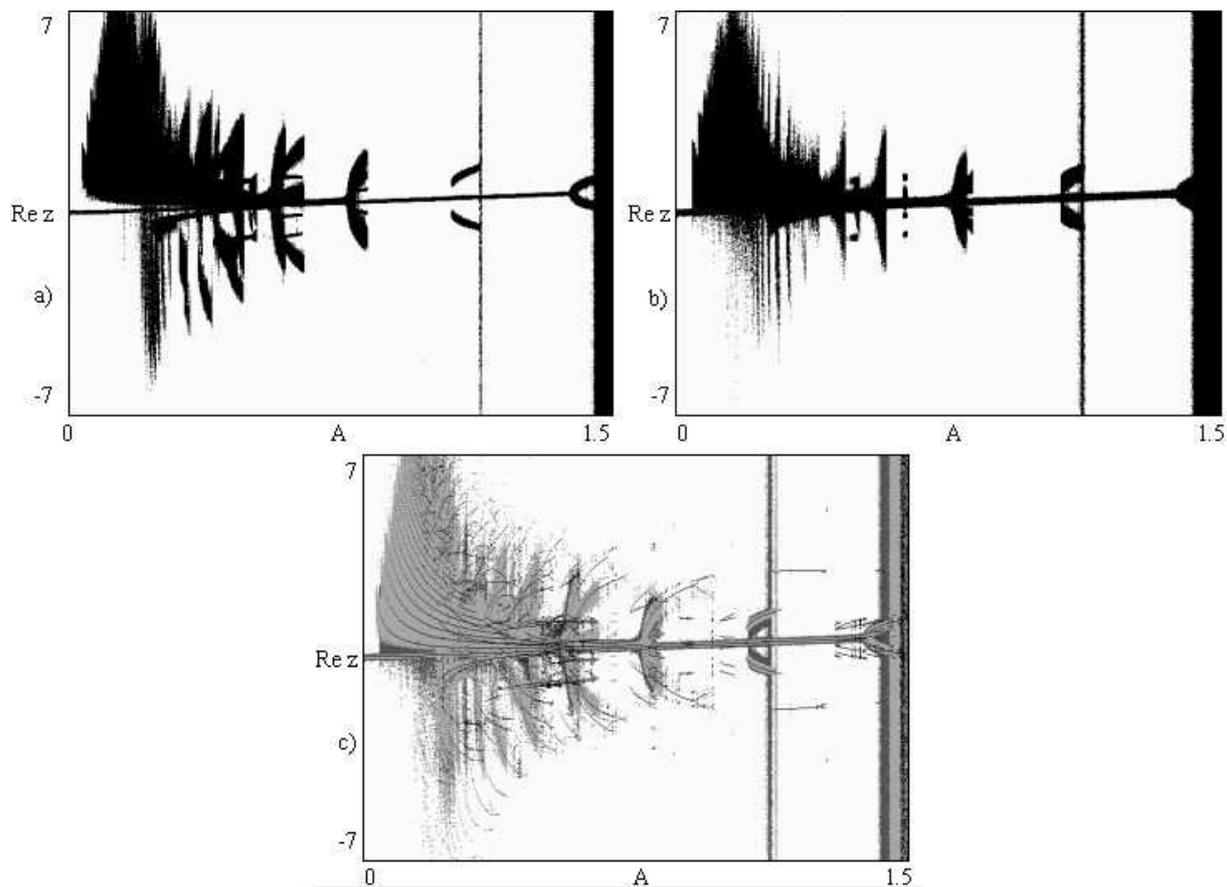}
\caption{Bifurcation diagrams for the map (4) for various values
of noise amplitude $\varepsilon$: a) $\varepsilon=0.005$; b)
$\varepsilon=0.01$; c) three diagrams are plotted over each other:
dark gray - $\varepsilon=0.005$; light gray - $\varepsilon=0.01$;
black - without noise. $B=0.99; \psi=0$. }
\end{figure}

Also it should be noted that some attractors undergo sharp
expansion before the disappearance so we can suppose that the
destruction of such attractors is a result of a collision of the
attractor with the boundary of its basin. Just before the
collision the attractor is very close to the basin boundary and it
seems likely that the trajectory can be thrown out of the basin by
noise which can lead to significant growth of the variable. The
realization of this dynamics confirms our suggestions that some
attractors undergo crisis.

\section{Conservative case of the Ikeda map}
Now let's present a brief analysis of the conservative case of the
Ikeda map. The Ikeda map (1) becomes the conservative system at
$B=1$. Plotting of the phase portraits is one of the basic methods
for its investigation. Phase portraits of the map (1) are
presented in the fig.~7. Their form is typical for driven
conservative nonlinear oscillator --- the families of invariant tori
corresponding to the existence of elliptic fixed points, some
"hollows" on them corresponding to the hyperbolic (saddle) fixed
points on the outside \cite{1} and periodic islands
surrounded by the domains of irregular dynamics, or the "chaotic
sea", exist. It can be clearly seen that at some parameter values the
"chaotic sea" exists not only outside, but also inside the
periodic islands. Furthermore, structures that are typical for
phase oscillations at nonlinear resonance \cite{1} can be seen on
the portraits.

For the investigation of the conservative system we propose a
method of plotting of so-called "divergence chart" that in some
sense is an analog to the chart of dynamical regimes for the
conservative systems. The procedure of its plotting is as follows.
For each point of $(A,\psi)$ plane we choose a set of points in the
phase space and fix the number of points that have stayed in the
finite region of phase space after a big number of iterations (we
use 15000 in numeric simulations). The different numbers of
non-diverged points correspond to the different shadows of gray
color. The comparison of "divergence chart" (fig. 8~b) and the
chart of dynamical regimes for nearly conservative case (fig. 8 a)
shows some correspondence between structures at the parameter
plane. For example, the border of the domain where practically all
points have gone to infinity in the conservative case corresponds
to the chaos border in dissipative system.

Now let's turn to the noisy driven conservative system (see (4)
with $B=1$). On the "divergence charts" (fig. 9) noise destroys
some small-scale structures, and the more noise amplitude is, the
more large-scale structures are destroyed.

On the phase portraits in noisy driven systems (fig. 10), as we
can predict, large-scale structures
\begin{figure}[h]
\includegraphics[width=15cm,keepaspectratio]{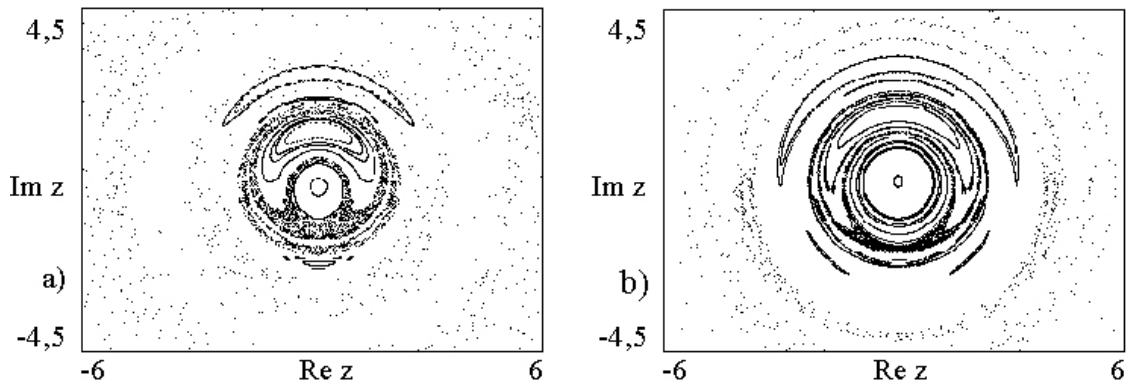}
\caption{Phase portraits of the map (1) in the conservative case
($B=1$). Parameters: a) $A=0.3, \psi=3\pi/2$; b) $A=0.2,
\psi=\pi$.}
\end{figure}
\begin{figure}[h]
\begin{centering}
\includegraphics[width=15cm,keepaspectratio]{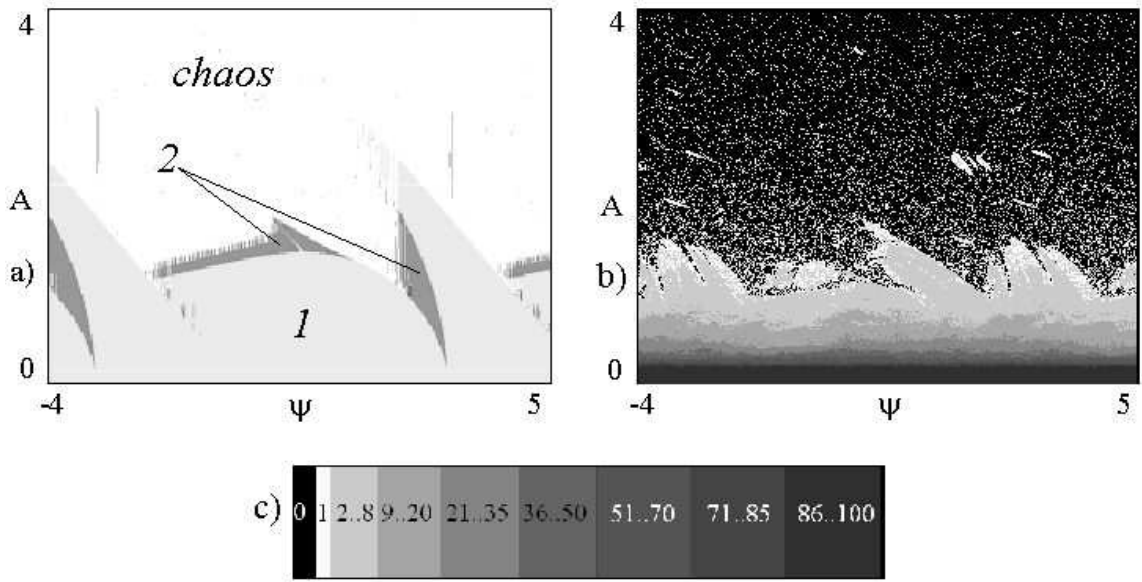}
\caption{Chart of dynamical regimes for dissipative map (1) (a,
$B=0.99$) and "divergence chart" for conservative map (1) (b,
$B=1$). Table of correspondence of colors to the numbers of points
that haven't gone to infinity is presented in fig.~c. In fig.~b
transient time is equal to 15000 iterations. }
\end{centering}
\end{figure}
\begin{figure}[p]
\includegraphics[width=16cm,keepaspectratio]{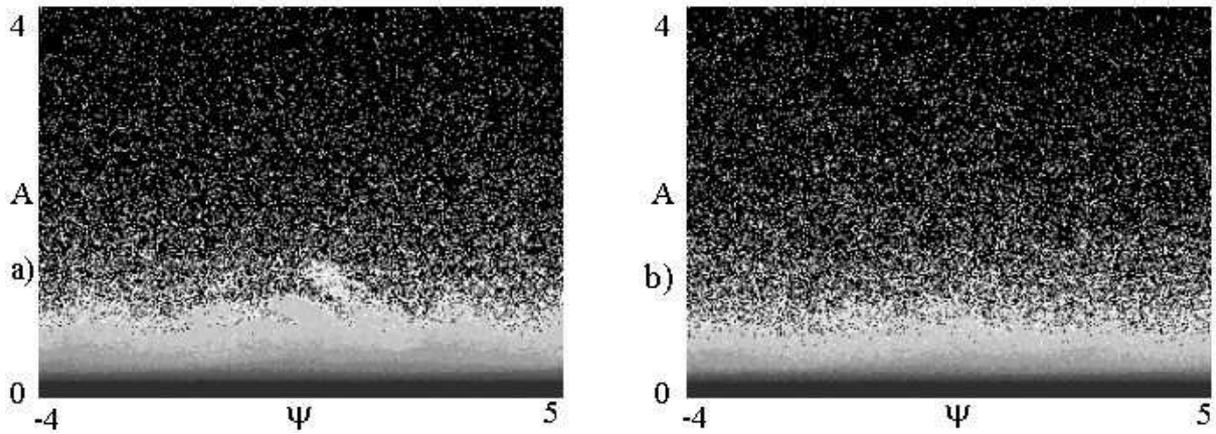}
\caption{"Divergence charts" for map (4) at different values of
noise amplitude: a) $\varepsilon=0.005$; b) $\varepsilon=0.01$.
Transient time is equal to 15000 iterations.}
\end{figure}
\begin{figure}[p]
\includegraphics[width=16cm,keepaspectratio]{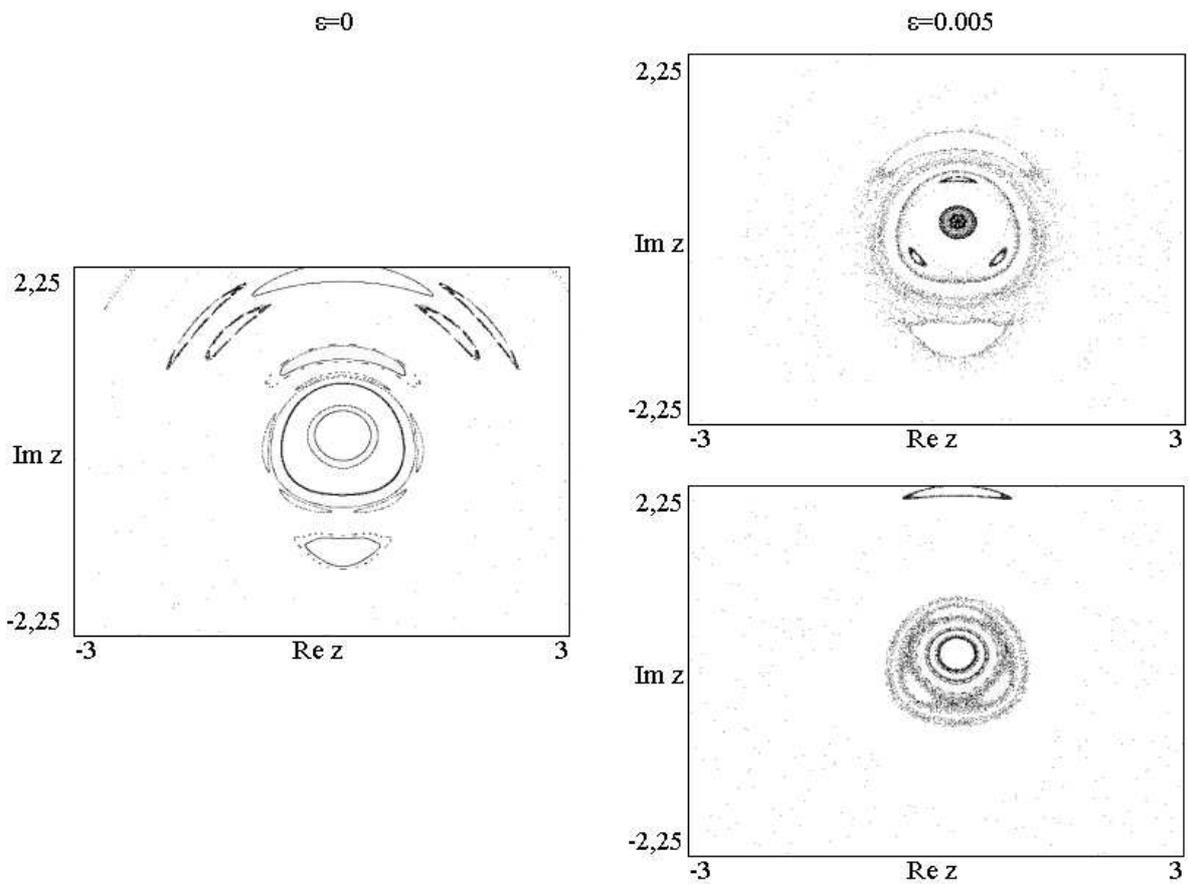}
\caption{Phase portrait for map (1) (a) and its different
realization for map (4) (b, c) with noise amplitude
$\varepsilon=0.005$. Parameters: $A=0.5; \psi=\pi/2; B=1$.}
\end{figure}
destroys a little but in general stay unchanged, and more small
structures destroy. It should be noted as a remarkable fact that
phase portraits of noisy driven system at fixed parameter values
can be significantly different (fig. 10 b, c). It can be explained
as follows. If a point in phase space lies near the separatrix
bounding two domains with different dynamics, it can be "thrown"
by noise influence from one domain to another so it will
demonstrate different dynamics on further stages of evolution. So
the more noise amplitude is the more wide is the band in which a
dynamics of the point can be changed.

\section{Conclusions}

Thus we have shown that the Ikeda map demonstrates a big number of
coexisting periodic attractors in the case of weak dissipation and
their number increase with the decreasing of dissipation. These
attractors can be divided into two types with different structure
and different length of the interval of the parameter A where they
exist.

The sharp increasing of transient time has been revealed with the
approaching of conservative case. At the beginning of the
transient process the system behavior is similar to conservative
and in the end to dissipative one. Moreover, it should be noted
that transient time depends essentially on the value of the
parameter A.

Also the sensitivity of the weakly dissipative system to the
external noise has been revealed: many of attractors are destroyed
by the noise of rather small amplitude. It can be explained as
follows. It is known that the noise effects the first stage of the
evolution much more then the stable regime, and the more
dissipative the system is, the faster it "forgets" initial
conditions. The system with very weak dissipation "remembers"
initial conditions for a very big time, hence, an external noise
influences on such systems more strongly.

At last, the new method for investigation the conservative case
was proposed. It was shown that structures similar to typical for
dissipative system arise at the parameter plane of conservative
system. Also it was shown that the conservative Ikeda system
demonstrates strong sensitivity to the noise influence.

{\it The work was supported by Russian Foundation for Basic Researches
(grant 04-02-04011 and 06-02-16773).}

\makeatletter
\renewcommand{\@biblabel}[1]{#1}
\makeatother


\begin{thebibliography}{xxxx}

\bibitem{1} G.~M.~Zaslavsky. {\it Physics of Chaos in Hamiltonian
Systems.} (Imperial College Press, 1998).


\bibitem{2} L.~E.~Reichl. {\it The Transition to Chaos in Conservative
Classical Systems: Quantum Manifes-tations.} (Springer-Verlag,
1992).


\bibitem{3} Feudel~U., Grebogi~C., Hunt~B.~R., Yorke~J.~A. Map with
more than 100 coexisting low-period periodic attractors. Physical
Review E. {\bf 54}, no. 1  71-81 (1996).


\bibitem{4} Ikeda~K., Daido~H., Akimoto~O.
Optical turbulence: Chaotic Behavior of Transmitted Light from a
Ring Cavity. Physical Review Letters. {\bf 45}, 709-712  (1980).


\bibitem{5} A.~P.~Kuznetsov, L.~V.~Turukina,
E.~Mosekilde. Dynamical systems of different classes as models of
the kicked nonlinear oscillator. International Journal of
Bifurcation and Chaos. {\bf 11}, no. 4  1065-1078  (2001).


\bibitem{6} A.~P.~Kuznetsov, S.~P.~Kuznetsov,
 E.~Mosekilde, L.~V.~Turukina. Two-parameter analysis of
the scaling behavior at the onset of chaos: tricritical and
pseudo-tricritical points. Physica A. {\bf 300}, 367-385 (2001).


\bibitem{7} E.~Mosekilde. {\it Topics in Nonlinear Dynamics.} (World Scientific, 1996).


\bibitem{8} Carcasses~J., Mira~C., Bosch~M., Simo~C., Tatjer~J.~C. Crossroad area - spring
area transition. (1) Parameter plane representation. International
Journal of Bifurcations and Chaos. {\bf 1}, 183-196  (1991).


\bibitem{9} Carcasses~J., Mira~C., Bosch~M., Simo~C. \& Tatjer~J.~C. "Crossroad area -
spring area transition" (II) foliated parametric representation.
International Journal of Bifurcations and Chaos. 1991, {\bf 1},
no. 2   339-348   (1991).


\bibitem{10}  Yu.~A.~Kuznetsov. {\it Elements of Applied Bifurcation
Theory.} (Springer-Verlag, 1998).


\end{thebibliography}
\end{document}